\documentstyle[multicol,floats,graphicx,subfigure,aps,pra]{revtex}
\begin{document}
\draft
\tighten
\small
\title{Complementarity and Young's interference fringes from two
atoms\cite{copyright}}
\author{ W.\ M.\ Itano, J.\ C.\ Bergquist, J.\ J.\ Bollinger,
D.\ J.\ Wineland}
\address{Time and Frequency Division, National Institute of Standards
           and Technology, Boulder, Colorado 80303}
\author{U.\ Eichmann}
\address{Max-Born-Institut for Non-linear Optics and Short Pulse
Spectroscopy, Rudower Chaussee 6, 12489 Berlin, Germany}
\author{M.\ G.\ Raizen}
\address{Department of Physics, University of Texas, Austin, Texas 78712}
\date{18 November 1997, Version 1.03}
\maketitle

\begin{abstract}
The interference pattern of the resonance fluorescence from a $J=1/2$
to $J=1/2$ transition of two identical atoms confined in a
three-dimensional harmonic potential is calculated.
Thermal motion of the atoms is included.
Agreement is obtained with experiments  [Eichmann {\em et al.},
Phys.\ Rev.\ Lett.\ {\bf 70}, 2359 (1993)].
Contrary to some theoretical predictions, but in agreement with the
present calculations, a fringe visibility greater than 50\% can be observed
with polarization-selective detection.
The dependence of the fringe visibility on polarization has a simple
interpretation, based on whether or not it is possible in principle to
determine which atom emitted the photon.
\end{abstract}

\begin{multicols}{2}
\section{Introduction}

Many variants of two-slit interference experiments, often ``thought
experiments,'' have been used to illustrate fundamental principles of quantum
mechanics.
Recently, Eichmann {\em et al.} \cite{eichmann93} have observed interference
fringes in the resonance fluorescence of two trapped ions,
analogous to those seen in Young's two-slit experiment.
Of particular interest was the fact that the interference fringes appeared
when it was impossible in principle to determine which ion which scattered
the photon disappeared when it was possible.
This is in agreement with Bohr's principle of complementarity,
which requires that the wave nature of the photon (the interference
fringes) cannot be observed under the same conditions as its particle
nature (the possibility of assigning to the photon a trajectory that
intersects just one of the ions).
In contrast to many thought experiments \cite{feynman65},
the disappearance of the fringes when the path of the particle can be
determined has nothing to do with the position-momentum indeterminacy
relations.
The experiment contains features from some thought experiments of
Scully and Dr\"uhl \cite{scully82}, regarding the interference of
light scattered by two multi-level atoms.

Recently, controversy has arisen over the mechanism by
which complementarity is enforced in a two-slit interference experiment.
Some claim that the destruction of interference by a determination
of the particle's path is {\em always} due to a random momentum transfer
necessitated by the indeterminacy relations
\cite{storey94,storey95,wiseman97}.
Others claim that the mere {\em existence} of the path information
can be sufficient to destroy the interference \cite{scully91}.
Englert {\em et al.} claim that the experiment of
Eichmann {\em et al.} supports the second position \cite{englert95}.

Published calculations explain some aspects of the
observations of Eichmann {\em et al.}
\cite{brewer95,brewer96,brewer96b,huang96,wong97}.
However, none of those calculations include all of the factors required
to make a comparison with the experimental data.
Here, we calculate the scattering cross section for arbitrary
directions and polarizations of the incident and outgoing light.
While the results were used in the analysis of the data in
Ref.~\cite{eichmann93},
the details of the calculations were not given.
The main limitation of the calculation is the use of perturbation
theory, so that it is valid only for low light intensities.
However, it includes the effect of thermal motion more precisely than
has been done elsewhere, taking into account the actual normal modes
of the system.
Also, the actual experimental geometry is fully taken into account, which
is not always the case in the other calculations.

Finally, we clarify the sense in which the loss of the fringe visibility
[defined as $(I_{\rm max}-I_{\rm min})/(I_{\rm max}+I_{\rm min})$]
for certain detected polarizations is due to the existence of ``which
path'' information in the ions.
This is an application of the fundamental quantum principle that
transition amplitudes are to be added before squaring if and only if
they connect the same initial and final states.

\section{Experiment}

\begin{figure}[htbp]
\begin{center}
\includegraphics[clip,width=2.5in]{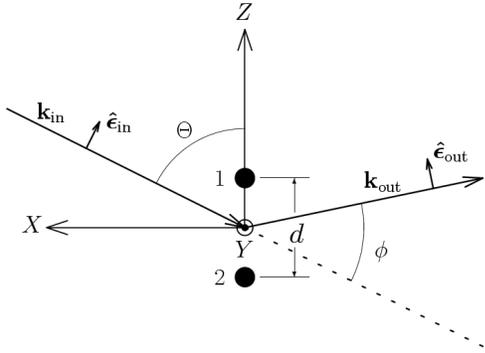}
\end{center}
\caption{Geometry of the Young's interference experiment, projected
onto the $X$-$Z$ plane.
The equilibrium positions of the two ions, represented by the filled circles,
lie along the $Z$ axis.
The wavevector ${\bf k}_{\rm in}$ of the incoming photon
is in the $X$-$Z$ plane, making an angle $\Theta$ with the $Z$ axis.
The $Y$ axis is out of the plane of the figure.
The projection of the wavevector ${\bf k}_{\rm out}$ onto the
$X$-$Z$ plane makes an angle $\phi$ with ${\bf k}_{\rm in}$.
The angle that ${\bf k}_{\rm out}$ deviates from the $X$-$Z$ plane
in the $+Y$ direction is $\Phi$ (not shown).
The polarization vectors of the incoming and outgoing photons are
$\bbox{\hat{\epsilon}}_{\rm in}$ and $\bbox{\hat{\epsilon}}_{\rm out}$.
}
\label{trap_coords}
\end{figure}

The experimental apparatus has been described previously
\cite{eichmann93,raizen92}.
Figure~\ref{trap_coords} shows the geometry.
Two $^{198}$Hg$^+$ ions were confined in a linear Paul (rf) trap
by a combination of static and rf electric fields.
The ions were laser-cooled to temperatures of a few mK
with a beam of linearly polarized, continuous-wave light, nearly resonant
with the 194 nm transition from the ground $6s\,^2S_{1/2}$ level to the
$6p\,^2P_{1/2}$ level.
The laser beam diameter was about 50 $\mu$m, and the power was 50 $\mu$W
or less.
The same beam was the coherent source for the Young's interference.
Cooling in the trap resulted in strong localization of the ions,
which was essential for observation of interference fringes.
The trap potentials were arranged so that a pair of ions would be oriented
along the symmetry ($Z$) axis of the trap.
The incoming photons, with wavevector ${\bf k}_{\rm in}$
and polarization vector $\bbox{\hat{\epsilon}}_{\rm in}$,
made an angle $\Theta$ of 62$^\circ$ with respect to the $Z$ axis.
The $X$ axis is oriented so that the $X$-$Z$ plane contains
${\bf k}_{\rm in}$.
Light emitted by the ions was collimated by a lens and directed to the
surface of an imaging photodetector, which was used to observe the fringes.
The wavevector and polarization of an outgoing photon are
${\bf k}_{\rm out}$ and $\bbox{\hat{\epsilon}}_{\rm out}$.
The projection of ${\bf k}_{\rm out}$ onto the $X$-$Z$ plane makes an
angle $\phi$ with respect to ${\bf k}_{\rm in}$.
The deviation of  ${\bf k}_{\rm out}$ from the $X$-$Z$ plane
in the $+Y$ direction is $\Phi$ (not shown in Fig.~\ref{trap_coords}).
The sensitive area of the photodetector included a range of $\phi$
from about 15$^\circ$ to 45$^\circ$ and a range of $\Phi$  from
about -15$^\circ$ to +15$^\circ$.
For polarization-selective detection, a glass plate oriented at Brewster's
angle was placed in the detection path so that nearly all of the light
with $\bbox{\hat{\epsilon}}_{\rm out}$ in the $X$-$Z$ plane was
transmitted into the glass, while some of the light polarized along
the $Y$ axis was reflected to the imaging detector.
The input polarization $\bbox{\hat\epsilon}_{\rm in}$ was varied.
Another lens system formed a real image of the ions on a second
imaging detector.
This image was used to determine when there were precisely
two ions in the trap.

\section{Two-ion harmonic oscillator system}

In the pseudo-potential approximation, the Hamiltonian for the
translational motion of the two ions in the harmonic trap is
\begin{equation}
H_{\rm trans}=\frac{{\bf P}_1^2}{2m}+\frac{{\bf P}_2^2}{2m}
+V({\bf R}_1)+V({\bf R}_2)
+\frac{e^2}{4\pi\epsilon_0\mid{\bf R}_1-{\bf R}_2\mid},
\label{h_trans}
\end{equation}
where ${\bf R}_i$ and ${\bf P}_i$ are the position and momentum of the
$i$th ion, $e$ and $m$ are the charge and mass of an ion, and
\begin{equation}
V({\bf R})\equiv\frac{1}{2}m\omega_R^2 \left(X^2+Y^2\right)
+\frac{1}{2}m\omega_Z^2 Z^2
\label{trap_potential}
\end{equation}
is the potential energy of a single ion in the trap.
In Eq.~(\ref{trap_potential}), we have made the approximation that
the trap pseudopotential is cylindrically symmetric.
Here, ${\bf R}$=$(X,Y,Z)$, in the Cartesian coordinate system shown
in Fig.~\ref{trap_coords}.
The classical equilibrium positions of the ions, found by minimizing the total
potential energy, are
${\bf R}_1^0=(d/2)\bbox{\hat{Z}}$ and ${\bf R}_2^0=-(d/2)\bbox{\hat{Z}}$,
where $d=[e^2/(2\pi\epsilon_0 m\omega_Z^2)]^{1/3}$,
and it is assumed that $\omega_R > \omega_Z$.

For small displacements ${\bf u}_1={\bf R}_1-{\bf R}_1^0$
and ${\bf u}_2={\bf R}_2-{\bf R}_2^0$
about the equilibrium positions, a harmonic approximation can be made.
The Hamiltonian separates into terms involving either center-of-mass (com)
or relative (rel) coordinates and momenta defined by
\begin{eqnarray}
{\bf u}^{\rm com}&\equiv&({\bf u}_1+{\bf u}_2)/2\nonumber\\
{\bf u}^{\rm rel}&\equiv&({\bf u}_1-{\bf u}_2)/2\nonumber\\
{\bf P}^{\rm com}&\equiv&{\bf P}_1+{\bf P}_2\\ \label{coord_trans}
{\bf P}^{\rm rel}&\equiv&{\bf P}_1-{\bf P}_2\nonumber
\end{eqnarray}
The translational Hamiltonian, in the harmonic approximation, is
\begin{eqnarray}
H_{\rm trans}&=&\hbar\omega_Z(N_Z^{\rm com}+1/2)
+\hbar\omega_R(N_X^{\rm com}+N_Y^{\rm com}+1)\nonumber\\
& &\mbox{}+\hbar\omega_S(N_Z^{\rm rel}+1/2)
+\hbar\omega_T(N_X^{\rm rel}+N_Y^{\rm rel}+1).
\label{h_trans2}
\end{eqnarray}
The number operators are defined in the usual way by
$N_i^{\rm com}\equiv(a_i^{\rm com})^\dagger
a_i^{\rm com}$ and $N_i^{\rm rel}\equiv(a_i^{\rm rel})^\dagger a_i^{\rm rel}$
for $i=X,Y,Z$.
The annihilation operators are defined in the usual way, for example:
\begin{equation}
a_Z^{\rm com}\equiv \sqrt{\frac{m\omega_Z}{\hbar}}u_Z^{\rm com}
+\frac{i}{\sqrt{4\hbar m\omega_Z}}P_Z^{\rm com}.
\label{a_defn}
\end{equation}
The three center-of-mass modes have the same frequencies as those of
a single ion in the trap: $\omega_Z$ and $\omega_R$.
The three relative modes include a symmetric stretch mode along the
$Z$ direction at frequency $\omega_S=\sqrt{3}\omega_Z$
and two tilting or rocking modes along the $X$ and $Y$ directions
at frequency $\omega_T=(\omega_R^2-\omega_Z^2)^{1/2}$
The eigenstates of $H_{\rm trans}$ are the simultaneous eigenstates
of the set of number operators
$\mid n_X^{\rm com}, n_Y^{\rm com}, n_Z^{\rm com},
n_X^{\rm rel}, n_Y^{\rm rel}, n_Z^{\rm rel}\rangle$ with eigenvalues
$\hbar[ \omega_Z(n_Z^{\rm com} +1/2)
+\omega_R(n_X^{\rm com}+n_Y^{\rm com}+1)
+\omega_S(n_Z^{\rm rel}+1/2)
+\omega_T(n_X^{\rm rel}+n_Y^{\rm rel}+1)]$.

\section{Atomic level structure}

\begin{figure}[htbp]
\begin{center}
\includegraphics[clip,width=2.5in]{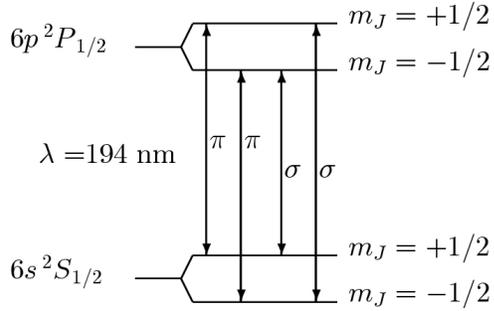}
\vspace{24pt}
\end{center}
\caption{Zeeman sublevels involved in the 194 nm,
6$s\,^2S_{1/2}$ to 6$p\,^2P_{1/2}$ transition of $^{198}$Hg$^+$.
The allowed $\pi$ and $\sigma$ transitions are labeled.
The Zeeman splitting of the levels is exaggerated.
}
\label{levelfig}
\end{figure}

Figure~\ref{levelfig} shows the magnetic sublevels involved in the
$6s\,^2S_{1/2}$ to $6p\,^2P_{1/2}$ transition.
These levels form an approximately closed system, since the
probability that the $6p\,^2P_{1/2}$ level radiatively decays to the
$5d^9 6s^2\,^2D_{3/2}$ level is only $1.4\times 10^{-7}$
\cite{itano87}.
The rest of the time it returns to the ground $6s\,^2S_{1/2}$ level.
The $5d^9 6s^2\,^2D_{3/2}$ level has a lifetime of 9 ms and decays
with about equal probability to the ground level or to the
$5d^9 6s^2\,^2D_{5/2}$, which has a lifetime of 86 ms and
decays only to the ground level.

Since the static magnetic field is small, we are free to define the
quantization axis of the ions to be along the electric polarization
vector $\bbox{\hat{\epsilon}}_{\rm in}$ of the incident light.
If the static magnetic field is along some other direction, then the Zeeman
sublevels defined according to the electric polarization vector
are not stationary states.
This does not change the analysis as long as the Zeeman precession
frequency is much less than the inverse of the scattering time,
which is approximately
equal to the 6$p\,^2P_{1/2}$-state lifetime (2.3 ns).
In the experiments described here, the magnetic field was
small enough that this was always the case.

\begin{figure}[htbp]
\begin{center}
\includegraphics[clip,width=3in]{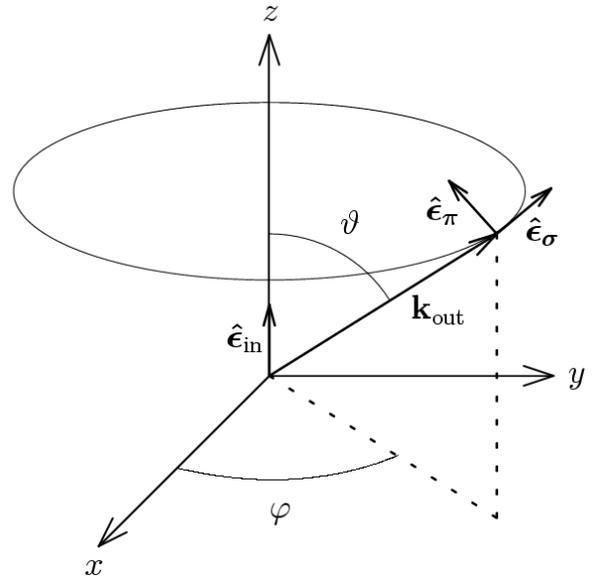}
\end{center}
\caption{Coordinate system for description of the direction and
polarization of the outgoing photon.
The $z$ axis is parallel to $\bbox{\hat{\epsilon}}_{\rm in}$,
and the $x$ axis is parallel to ${\bf k}_{\rm in}$.
The polarization vector $\bbox{\hat{\epsilon}_{\pi}}$ lies in the plane
containing $\bbox{\hat{\epsilon}}_{\rm in}$ and ${\bf k}_{\rm out}$,
while $\bbox{\hat{\epsilon}_{\sigma}}$ is perpendicular to that plane.
}
\label{atom_coords}
\end{figure}

Figure \ref{atom_coords} shows a Cartesian coordinate system having
its $z$ axis oriented along $\bbox{\hat{\epsilon}}_{\rm in}$.
The $x$ axis is parallel to ${\bf k}_{\rm in}$.
The $y$ axis is defined so that $(x,y,z)$ forms a right-handed coordinate
system.
This coordinate system is more useful than the trap-oriented $(X,Y,Z)$
coordinate system of Fig.~\ref{trap_coords} for
describing the angular distribution of the scattered light.

\section{Scattering cross section}

Consider the process in which two ions, initially in their ground
electronic states, absorb a photon having wavevector ${\bf k_{\rm in}}$
and polarization $\bbox{\hat\epsilon}_{\rm in}$,
emit a photon having wavevector ${\bf k_{\rm out}}$ and polarization
$\bbox{\hat\epsilon}_{\rm out}$, and are left in their ground electronic states.
The ions may change their Zeeman sublevels during the process.
Also, the motional state of the two ions may change.

The electric dipole Hamiltonian which causes the transitions is
\begin{equation}
H_{\rm ED}=-{\bf D}_1\cdot{\bf E}({\bf R}_1, t)
-{\bf D}_2\cdot{\bf E}({\bf R}_2, t),
\label{electric-dipole}
\end{equation}
where ${\bf D}_1$ and ${\bf D}_2$ are the electric dipole moment operators
for ions 1 and 2 and ${\bf E}({\bf R}, t)$ is the electric field, consisting
of a classical part, representing the incident laser beam, and the
quantized free field operator:
\begin{eqnarray}
{\bf E}({\bf R}, t)&=&\bbox{\hat{\epsilon}}_{\rm in}\Re {\cal E}_0
e^{i{\bf k}_{\rm in}\cdot{\bf R}
-i\omega_{\rm in}t}\nonumber\\
& & \mbox{}+\sum_s i\sqrt{\frac{\hbar \omega_s}{2\epsilon_0 V}}
\left[a_s\bbox{\hat{\epsilon}}_s e^{i{\bf k}_s\cdot {\bf R}}
-a_s^\dagger\bbox{\hat{\epsilon}}_s e^{-i{\bf k}_s\cdot {\bf R}}\right],
\label{e-field}
\end{eqnarray}
where $\Re$ denotes the real part, ${\cal E}_0$ is the amplitude
of the laser electric field, and $a_s$ is the annihilation operator
for a photon of wavevector ${\bf k}_s$, frequency $\omega_s$, and
polarization $\bbox{\epsilon}_s$,
and $V$ is the quantization volume.

The electric dipole Hamiltonian,  in second-order perturbation
theory, gives the cross section for the two ions to scatter a photon
in a particular direction:
\end{multicols}
\vspace{6pt}
\begin{eqnarray}
\frac{d\sigma_i}{d\Omega_{\rm out}}
&=&\sum_f C_1\left|\sum_j\frac{\langle\Psi_f|({\bf D}_1\cdot
\bbox{\hat\epsilon}_{\rm out})
e^{-i{\bf k_{\rm out}\cdot R}_1}
|\Psi_j\rangle\langle\Psi_j|({\bf D}_1\cdot\bbox{\hat\epsilon}_{\rm in})
e^{i{\bf k_{\rm in}\cdot R}_1}|\Psi_i\rangle}
{\omega_0-\omega_{\rm in}+(E_j-E_i)/\hbar-i\gamma/2}
\right.\nonumber\\
& &+\left.\sum_j\frac{\langle\Psi_f|({\bf D}_2\cdot
\bbox{\hat\epsilon}_{\rm out}) e^{-i{\bf k_{\rm out}\cdot R}_2}
|\Psi_j\rangle\langle\Psi_j|({\bf D}_2\cdot\bbox{\hat\epsilon}_{\rm in})
e^{i{\bf k_{\rm in}\cdot R}_2}|\Psi_i\rangle}
{\omega_0-\omega_{\rm in}+(E_j-E_i)/\hbar-i\gamma/2}
\right|^2,
\label{crosssection1}
\end{eqnarray}
\begin{multicols}{2}
\noindent where $\omega_{\rm in}=c|{\bf k}_{\rm in}|$,
$\omega_{\rm out}=c|{\bf k}_{\rm out}|$, $\hbar\omega_0$ is the
separation between the ground and excited electronic states of an ion,
$\gamma$ is the decay rate of the excited state, and
$C_1\equiv \omega_{\rm out}^3/(16\pi^2c^4\hbar^2\epsilon_0^2)$.
The initial, final, and intermediate states describing the electronic
and motional degrees of freedom of the system are
$|\Psi_i\rangle$, $|\Psi_f\rangle$, and $|\Psi_j\rangle$.
The energies $E_i$, $E_f$, and $E_j$ are the motional energies of the ions
in the initial, final, and intermediate states.
They depend on the values of the six harmonic oscillator quantum
numbers, which we denote by $\{n_{\rm ho}\}$.
Because of energy conservation, the frequency of the outgoing photon depends
on the final state:
\begin{equation}
\omega_{\rm out}=\omega_{\rm in}+(E_i-E_f)/\hbar.
\end{equation}
Thus, the scattered light has a discrete frequency spectrum, and the
different components could, in principle, be detected separately.
In Eq.~(\ref{crosssection1}), all frequency components are summed,
which is appropriate if
the detection is frequency-insensitive.
The laser frequency is assumed to be nearly resonant
with an optical transition in the ion, so that only one intermediate
electronic state has to be included in the sums,  and we can neglect
the counter-rotating terms.
We ignore dipole-dipole interactions between the ions, because they
were separated by many wavelengths in the experiment.
Here we specialize to the case of an ion which has no nuclear
spin and which has a $^2S_{1/2}$ ground state and a $^2P_{1/2}$
excited state,
like the $^{198}$Hg$^+$ ions used in Ref.~\cite{eichmann93}.
We denote a state in which ion 1 is in the $(^2S_{1/2}$, $m_J=+1/2)$ state,
ion 2 is in the $(^2P_{1/2}$, $m_J=-1/2)$ state,
and has the harmonic oscillator quantum numbers $\{n_{\rm ho}\}$ by
\begin{equation}
|\Psi\rangle=|(^2S_{1/2},+1/2)_1 (^2P_{1/2},-1/2)_2
\{n_{\rm ho}\}\rangle.
\end{equation}

There are four possible sets of initial $m_J$ quantum numbers for the two
ions and four possible final sets.
There are two basic kinds of scattering processes --- those which preserve
the $m_J$ quantum numbers of the ions, and those which
change $m_J$ of one ion.
We treat these cases separately.
The form of Eq.~(\ref{crosssection1}) excludes the possibility
of both ions changing their $m_J$ quantum numbers.

\subsection{Both $\bbox{m_J}$ quantum numbers remain the same
($\bbox{\pi}$ case)}

In order to be definite, we let $m_J=+1/2$ for both ions,
both before and after the scattering.
That is,
\begin{eqnarray}
|\Psi_i\rangle& = &|(^2S_{1/2},+1/2)_1 (^2S_{1/2},+1/2)_2
\{n_{\rm ho}\}_i\rangle,\\
|\Psi_f\rangle& = &|(^2S_{1/2},+1/2)_1 (^2S_{1/2},+1/2)_2
\{n_{\rm ho}\}_f\rangle.
\end{eqnarray}
We call this the $\pi$ case, because it involves only $\pi$ transitions,
that is, transitions that leave $m_J$ unchanged.
Because of the electric dipole selection rules, the only intermediate
states which contribute nonzero terms are of the form
\begin{equation}
|\Psi_j\rangle= |(^2P_{1/2},+1/2)_1 (^2S_{1/2},+1/2)_2
\{n_{\rm ho}\}_j\rangle
\end{equation}
for the first sum, and
\begin{equation}
|\Psi_j\rangle= |(^2S_{1/2},+1/2)_1 (^2P_{1/2},+1/2)_2
\{n_{\rm ho}\}_j\rangle
\end{equation}
for the second sum.
The matrix elements connecting the initial states to the intermediate
states are
\begin{eqnarray}
\lefteqn{\langle\Psi_j|({\bf D}_p\cdot\bbox{\hat\epsilon}_{\rm in})
e^{i{\bf k_{\rm in}\cdot R}_p}|\Psi_i\rangle}\nonumber\\
&=&\langle(^2P_{1/2},+1/2)_p|D_{pz}|(^2S_{1/2},+1/2)_p\rangle\nonumber\\
& &\times\langle\{n_{\rm ho}\}_j|e^{i{\bf k_{\rm in}\cdot R}_p}|
\{n_{\rm ho}\}_i\rangle\nonumber\\
&=&\frac{1}{\sqrt{6}}(^2P_{1/2}\|D^{(1)}\|^2S_{1/2})
\langle\{n_{\rm ho}\}_j|e^{i{\bf k_{\rm in}\cdot R}_p}|
\{n_{\rm ho}\}_i\rangle,
\end{eqnarray}
where $p$ = 1 or 2, $D_{pz}$ is the $z$ component of the
${\bf D}_p$ operator, and $(^2P_{1/2}\|D^{(1)}\|^2S_{1/2})$
is the reduced matrix element of the dipole moment operator
(the same for both ions).

The angular distribution of the outgoing photon is contained
in the matrix elements connecting the intermediate states to the
final states.
The unit propagation vector for the outgoing photon is
\begin{equation}
\bbox{\hat{\rm k}}_{\rm out}=
(\sin\vartheta\,\cos\varphi,\sin\vartheta\,\sin\varphi,\cos\vartheta),
\label{kout}
\end{equation}
where $\vartheta$ and $\varphi$ are spherical polar angles with respect
to the $(x,y,z)$ coordinate system of Fig.~\ref{atom_coords}.
The polarization vector $\bbox{\hat{\epsilon}}_{\rm out}$ must be perpendicular to
$\bbox{\hat{\rm k}}_{\rm out}$.
We define two mutually orthogonal unit polarization vectors,
both perpendicular to $\bbox{\hat{\rm k}}_{\rm out}$, by
\begin{equation}
\bbox{\hat\epsilon}_{\pi}=
(-\cos\vartheta\,\cos\varphi,-\cos\vartheta\,\sin\varphi,\sin\vartheta)
\end{equation}
and
\begin{equation}
\bbox{\hat{\epsilon}}_{\sigma}=(-\sin\varphi,\cos\varphi,0).
\end{equation}
Since only the $z$ components of ${\bf D}_1$ and ${\bf D}_2$
contribute to the matrix elements connecting the intermediate
and final states for this case, light with polarization vector
$\bbox{\hat{\epsilon}}_{\sigma}$ cannot be emitted.

With the choice of $\bbox{\hat\epsilon}_{\rm out}$=$\bbox{\hat\epsilon}_{\pi}$,
the matrix elements connecting the intermediate states to the final
states are

\begin{eqnarray}
\lefteqn{\langle\Psi_f|({\bf D}_p\cdot\bbox{\hat\epsilon}_{\pi})
e^{-i{\bf k_{\rm out}\cdot R}_p}|\Psi_j\rangle}\nonumber\\
&=&\sin\vartheta\langle(^2S_{1/2},+1/2)_p|D_{pz}|(^2P_{1/2},+1/2)_p\rangle
\nonumber\\
& &\times
\langle\{n_{\rm ho}\}_f|e^{-i{\bf k_{\rm out}\cdot R}_p}|
\{n_{\rm ho}\}_j\rangle\nonumber\\
&=&\frac{\sin\vartheta}{\sqrt{6}}(^2S_{1/2}\|D^{(1)}\|^2P_{1/2})
\langle\{n_{\rm ho}\}_f|e^{-i{\bf k_{\rm out}\cdot R}_p}|
\{n_{\rm ho}\}_j\rangle.
\end{eqnarray}
\end{multicols}

Equation~(\ref{crosssection1}) for the cross section becomes
\begin{eqnarray}
\frac{d\sigma^{(1)}}{d\Omega_{\rm out}}
&=&\frac{\sin^2\vartheta}{36}|(^2S_{1/2}\|D^{(1)}\|^2P_{1/2})|^4
\sum_f C_1\left|\sum_j\frac{\langle\{n_{\rm ho}\}_f|
e^{-i{\bf k_{\rm out}\cdot R}_1}|\{n_{\rm ho}\}_j\rangle
\langle\{n_{\rm ho}\}_j|
e^{i{\bf k_{\rm in}\cdot R}_1}|\{n_{\rm ho}\}_i\rangle}
{\omega_0-\omega_{\rm in}+(E_j-E_i)/\hbar-i\gamma/2}
\right.\nonumber\\
& &+\left.\sum_j\frac{\langle\{n_{\rm ho}\}_f|
e^{-i{\bf k_{\rm out}\cdot R}_2}
|\{n_{\rm ho}\}_j\rangle
\langle\{n_{\rm ho}\}_j|
e^{i{\bf k_{\rm in}\cdot R}_2}|\{n_{\rm ho}\}_i\rangle}
{\omega_0-\omega_{\rm in}+(E_j-E_i)/\hbar-i\gamma/2}
\right|^2.
\label{crosssection2}
\end{eqnarray}

\begin{multicols}{2}
\noindent The $(1)$ superscript on the cross section is to label it as
the $\pi$ case.
The same result would have
been obtained for any of the other three possible sets of initial
$m_J$ quantum numbers, so this is the general result for the
case in which the $m_J$ values do not change.
The presence of two terms in Eq.~(\ref{crosssection2}), which are
added and then squared, is the source of the Young's interference fringes.
These two terms can be identified with the two possible
paths for the photon, each intersecting one of the two ions.

The sums over intermediate harmonic oscillator states
can be done by closure if the energy denominators are constant.
While they are not constant, because $(E_j-E_i)$ varies,
it can be shown (see, for example, Ref.~\cite{lipkin62})
that the main contributions to the sum come from terms where
$|E_j-E_i|$ is less than or on the order of $\sqrt{RE_i}$, where $R$
is the photon recoil energy $(\hbar k_{\rm out})^2/(2m)$.
For the Hg$^+$ 194.2 nm transition, $R$=$h\times 26.7$ kHz.
For Doppler cooling, $E_i$ will be on the order of $\hbar\gamma$,
where, for this transition, $\hbar\gamma$=$h\times 70$ MHz.
Thus, the rms value of $(E_j-E_i)$ will be on the order of
$h\times1.4$ MHz, which is much less than $\hbar\gamma/2$.
Therefore, while the denominators are not strictly constant, they are
nearly constant for the terms which contribute significantly to the
sums.

If we neglect $(E_j-E_i)/\hbar$ compared to $\gamma/2$
and use closure to evaluate the sums,
Eq.~(\ref{crosssection2}) simplifies to
\end{multicols}

\vspace{6pt}
\begin{eqnarray}
\frac{d\sigma^{(1)}}{d\Omega_{\rm out}}
&=&\frac{\sin^2\vartheta}{36}
\frac{|(^2S_{1/2}\|D^{(1)}\|^2P_{1/2})|^4}
{(\omega_0-\omega_{\rm in})^2+\gamma^2/4}
\sum_f C_1\left|\langle\{n_{\rm ho}\}_f|
e^{-i{\bf k_{\rm out}\cdot R}_1}
e^{i{\bf k_{\rm in}\cdot R}_1}|\{n_{\rm ho}\}_i\rangle
+\langle\{n_{\rm ho}\}_f|
e^{-i{\bf k_{\rm out}\cdot R}_2}
e^{i{\bf k_{\rm in}\cdot R}_2}|\{n_{\rm ho}\}_i\rangle
\right|^2\nonumber\\
&=&\frac{\sin^2\vartheta}{36}
\frac{|(^2S_{1/2}\|D^{(1)}\|^2P_{1/2})|^4}
{(\omega_0-\omega_{\rm in})^2+\gamma^2/4}
\sum_f C_1\left|\langle\{n_{\rm ho}\}_f|
e^{-i{\bf q\cdot R}_1}
|\{n_{\rm ho}\}_i\rangle
+\langle\{n_{\rm ho}\}_f|
e^{-i{\bf q\cdot R}_2}
|\{n_{\rm ho}\}_i\rangle
\right|^2\nonumber\\
&=&\frac{\sin^2\vartheta}{36}
\frac{|(^2S_{1/2}\|D^{(1)}\|^2P_{1/2})|^4}
{(\omega_0-\omega_{\rm in})^2+\gamma^2/4}
\sum_f C_1\left|\langle\{n_{\rm ho}\}_f|
e^{-i{\bf q\cdot R}_1}+e^{-i{\bf q\cdot R}_2}
|\{n_{\rm ho}\}_i\rangle
\right|^2,
\label{crosssection3}
\end{eqnarray}
\begin{multicols}{2}
\noindent where ${\bf q}\equiv{\bf k}_{\rm out}-{\bf k}_{\rm in}$.
Since the branching ratio for decay of the excited $^2P_{1/2}$ states
to the ground $^2S_{1/2}$ states is nearly 100\%, the
spontaneous decay rate $\gamma$ is
\begin{equation}
\gamma=\frac{\omega_0^3}{6\pi\epsilon_0\hbar c^3}\left|
(^2S_{1/2}\|D^{(1)}\|^2P_{1/2})\right|^2.
\end{equation}

Equation (\ref{crosssection3}) for the cross section becomes
\begin{eqnarray}
\frac{d\sigma^{(1)}}{d\Omega_{\rm out}}
&=&\frac{\sin^2 \vartheta}{8\pi}\sigma_0{\cal L}(\omega_{\rm in}-\omega_0)
\nonumber\\
& &\times\sum_f \left|\langle\{n_{\rm ho}\}_f|
e^{-i{\bf q\cdot R}_1}+e^{-i{\bf q\cdot R}_2}
|\{n_{\rm ho}\}_i\rangle\right|^2,
\label{crosssection4}
\end{eqnarray}
where
$\sigma_0=\lambda_0^2/2\pi$ is
the resonance cross section, $\lambda_0=2\pi c/\omega_0$ is the
resonance wavelength,
and ${\cal L}(\omega_{\rm in}-\omega_0)$
is a Lorentzian of unit height and width $\gamma$:
\begin{equation}
{\cal L}(\omega_{\rm in}-\omega_0)\equiv
\frac{(\gamma/2)^2}{(\omega_{\rm in}-\omega_0)^2+(\gamma/2)^2}.
\end{equation}
In deriving Eq.~(\ref{crosssection4}), we have assumed that
$\omega_0/\omega_{\rm out}\approx 1$.
The sum over final harmonic oscillator states can be done by closure:
\end{multicols}

\vspace{12pt}
\begin{eqnarray}
\frac{d\sigma^{(1)}}{d\Omega_{\rm out}}
&=&\frac{\sin^2 \vartheta}{8\pi}\sigma_0{\cal L}(\omega_{\rm in}-\omega_0)
\langle\{n_{\rm ho}\}_i|
(e^{i{\bf q\cdot R}_1}+e^{i{\bf q\cdot R}_2})
(e^{-i{\bf q\cdot R}_1}+e^{-i{\bf q\cdot R}_2})
|\{n_{\rm ho}\}_i\rangle\nonumber\\
&=&\frac{\sin^2 \vartheta}{8\pi}\sigma_0{\cal L}(\omega_{\rm in}-\omega_0)
\langle\{n_{\rm ho}\}_i| 2+
e^{i{\bf q}\cdot ({\bf R}_1-{\bf R}_2)}+
e^{-i{\bf q}\cdot ({\bf R}_1-{\bf R}_2)}
|\{n_{\rm ho}\}_i\rangle.
\label{crosssection5}
\end{eqnarray}
The exponentials can be combined in Eq.~(\ref{crosssection5})
because the components of ${\bf R}_1$ and ${\bf R}_2$ commute.
The cross section can be written in terms of the equilibrium ion
separation ${\bf d}$ and
the displacement coordinates ${\bf u}_1$ and ${\bf u}_2$ as
\begin{equation}
\frac{d\sigma^{(1)}}{d\Omega_{\rm out}}
=\frac{\sin^2 \vartheta}{8\pi}\sigma_0{\cal L}(\omega_{\rm in}-\omega_0)
\langle\{n_{\rm ho}\}_i| 2+
e^{i{\bf q}\cdot ({\bf d}+{\bf u}_1-{\bf u}_2)}+
e^{-i{\bf q}\cdot ({\bf d}+{\bf u}_1-{\bf u}_2)}
|\{n_{\rm ho}\}_i\rangle.
\label{crosssection6}
\end{equation}
The exponential factors in Eq.~(\ref{crosssection6}) depend on the
relative coordinates of the two ions and not on their center-of-mass
coordinates.                           

In order to compare with the experiment, we compute the cross section
averaged over a thermal distribution of
$|\{n_{\rm ho}\}_i\rangle$ initial states:
\begin{equation}
\left\langle\frac{d\sigma^{(1)}}{d\Omega_{\rm out}}\right\rangle
=
\frac{\sin^2 \vartheta}{8\pi}\sigma_0{\cal L}(\omega_{\rm in}-\omega_0)
\left[ 2+ e^{i{\bf q}\cdot{\bf d}}
\left\langle e^{i{\bf q}\cdot({\bf u}_1-{\bf u}_2)}\right\rangle
+e^{-i{\bf q}\cdot {\bf d}}
\left\langle e^{-i{\bf q}\cdot({\bf u}_1-{\bf u}_2)}\right\rangle
\right]
\label{crosssection7}
\end{equation}
where $\langle A \rangle$ denotes the thermal average of the
operator $A$.
For harmonic oscillators, the thermal averages have a simple form
\cite{mermin66,bateman92}:
\begin{equation}
\langle e^{\pm i{\bf q}\cdot({\bf u}_1-{\bf u}_2)}\rangle
=e^{-\frac{1}{2}\langle[{\bf q}\cdot({\bf u}_1-{\bf u}_2)]^2\rangle}.
\label{thermal-avg}
\end{equation}
While Refs.~\cite{mermin66,bateman92}
assume a common temperature for all of the
harmonic oscillator modes, Eq.~(\ref{thermal-avg}) is still valid
if different modes have different temperatures.
Different modes are laser-cooled at different rates depending
on the direction of the laser beam.
Hence, the modes have different temperatures unless the energy transfer rate
between them is fast \cite{itano82}.
The thermally averaged cross section is
\begin{equation}
\left\langle\frac{d\sigma^{(1)}}{d\Omega_{\rm out}}\right\rangle
=
\frac{\sin^2\vartheta}{4\pi}\sigma_0{\cal L}(\omega_{\rm in}-\omega_0)
\left[1+\cos({\bf q\cdot d})
e^{-\frac{1}{2}\langle[{\bf q}\cdot({\bf u}_1-{\bf u}_2)]^2\rangle}
\right],
\label{crosssection10}
\end{equation}
which is equivalent to Eq.~(1) of Ref.~\cite{eichmann93},
except that it includes the $\sin^2\vartheta$ angular dependence.
The interference fringe visibility is given by the exponential
factor multiplying $\cos({\bf q}\cdot{\bf d})$
This factor decreases with increasing temperature and
is analogous to the
Debye-Waller factor for x ray scattering from a crystal.
It can be rewritten as
\begin{eqnarray}
e^{-\frac{1}{2}\langle[{\bf q}\cdot({\bf u}_1-{\bf u}_2)]^2\rangle}
&=&\exp\left[
-\frac{\hbar q_X^2}{m\omega_T}
\left(\langle N_X^{\rm rel}\rangle + \frac{1}{2}\right)
-\frac{\hbar q_Y^2}{m\omega_T}
\left(\langle N_Y^{\rm rel}\rangle + \frac{1}{2}\right)
-\frac{\hbar q_Z^2}{m\omega_S}
\left(\langle N_Z^{\rm rel}\rangle + \frac{1}{2}\right)
\right]\nonumber\\
&=&\exp\left[
-\frac{\hbar q_X^2}{2m\omega_T}\coth
\left(\frac{\hbar\omega_T}{2k_BT_X^{\rm rel}}\right)
-\frac{\hbar q_Y^2}{2m\omega_T}\coth
\left(\frac{\hbar\omega_T}{2k_BT_Y^{\rm rel}}\right)
-\frac{\hbar q_Z^2}{2m\omega_S}\coth
\left(\frac{\hbar\omega_S}{2k_BT_Z^{\rm rel}}\right)
\right]\nonumber\\
&\approx&\exp\left(-\frac{q_X^2 k_B T_X^{\rm rel}}{m\omega_T^2}
-\frac{q_Y^2 k_B T_Y^{\rm rel}}{m\omega_T^2}
-\frac{q_Z^2 k_B T_Z^{\rm rel}}{m\omega_S^2}
\right),
\label{debye-waller}
\end{eqnarray}
\begin{multicols}{2}
\noindent
where $T_Z^{\rm rel}$ is the temperature of the $u_Z^{\rm rel}$ mode, etc.,
and the approximation in the last line is valid when the
mean harmonic oscillator quantum numbers are large.
In the limit of small thermal motion or small $|{\bf q}|$
(near-forward scattering), the visibility
can approach 100\% (with polarized detection),
in agreement with Ref.~\cite{wong97},
but in contradiction to Ref.~\cite{brewer96b},
where it was claimed that the visibility could not exceed 50\%.

\subsection{One $\bbox{m_J}$ quantum number changes ($\bbox{\sigma}$ case)}

Here we consider the case in which one of the ions changes its
$m_J$ quantum number in the scattering process.
We call this the $\sigma$ case, since it involves a $\sigma$ transition,
that is, a transition that changes $m_J$ by $\pm 1$ in one of the ions.
There are eight cases, since there are four possible initial states
and two ions which could change quantum numbers.

In order to be definite, we pick the case where $m_J=+1/2$ for both ions
before the scattering, and ion 1 changes to $m_J=-1/2$ after the scattering.
That is,
\begin{eqnarray}
|\Psi_i\rangle& = &|(^2S_{1/2},+1/2)_1 (^2S_{1/2},+1/2)_2
\{n_{\rm ho}\}_i\rangle,\\
|\Psi_f\rangle& = &|(^2S_{1/2},-1/2)_1 (^2S_{1/2},+1/2)_2
\{n_{\rm ho}\}_f\rangle.
\end{eqnarray}
Only the first sum over $j$ in Eq.~(\ref{crosssection1})
contributes, since only it contains ${\bf D}_1$, the dipole
moment which leads to the change in $m_J$ of ion 1.

As in the previous case, the only intermediate
states which contribute nonzero terms are of the form
\begin{equation}
|\Psi_j\rangle= |(^2P_{1/2},+1/2)_1 (^2S_{1/2},+1/2)_2
\{n_{\rm ho}\}_j\rangle.
\end{equation}
The matrix elements connecting the initial states to the intermediate
states are
\begin{eqnarray}
\lefteqn{\langle\Psi_j|({\bf D}_1\cdot\bbox{\hat\epsilon}_{\rm in})
e^{i{\bf k}_{\rm in}\cdot{\bf R}_1}|\Psi_i\rangle}\nonumber\\
&=&\langle(^2P_{1/2},+1/2)_1|D_{1z}|(^2S_{1/2},+1/2)_1\rangle\nonumber\\
& &\times\langle\{n_{\rm ho}\}_j|e^{i{\bf k}_{\rm in}\cdot {\bf R}_1}|
\{n_{\rm ho}\}_i\rangle\nonumber\\
&=&\frac{1}{\sqrt{6}}(^2P_{1/2}\|D^{(1)}\|^2S_{1/2})
\langle\{n_{\rm ho}\}_j|e^{i{\bf k}_{\rm in}\cdot {\bf R}_1}|
\{n_{\rm ho}\}_i\rangle.
\end{eqnarray}

In the $\vartheta=\pi/2$ plane, only the polarization corresponding to
$\bbox{\hat\epsilon}_{\sigma}$ is emitted, but in general,
light with both $\bbox{\hat\epsilon}_{\sigma}$
and $\bbox{\hat\epsilon}_{\pi}$ contributes to the scattered intensity.
We consider these two cases separately.

For $\bbox{\hat\epsilon}_{\rm out}$=$\bbox{\hat\epsilon}_{\sigma}$,
the matrix elements connecting the intermediate states to the final
states are
\begin{eqnarray}
\lefteqn{\langle\Psi_f|({\bf D}_1\cdot\bbox{\hat\epsilon}_{\sigma})
e^{-i{\bf k}_{\rm out}\cdot {\bf R}_1}|\Psi_j\rangle}\nonumber\\
&=&\frac{-ie^{i\varphi}}{\sqrt{2}}
\langle(^2S_{1/2},-1/2)_1|D_{1 -1}^{(1)}|(^2P_{1/2},+1/2)_1\rangle\nonumber\\
& &\times\langle\{n_{\rm ho}\}_f|e^{-i{\bf k_{\rm out}\cdot R}_1}|
\{n_{\rm ho}\}_j\rangle\\
&=&\frac{-ie^{i\varphi}}{\sqrt{6}}(^2S_{1/2}\|D^{(1)}\|^2P_{1/2})
\langle\{n_{\rm ho}\}_f|e^{-i{\bf k}_{\rm out}\cdot {\bf R}_1}|
\{n_{\rm ho}\}_j\rangle,\nonumber
\end{eqnarray}
where   $D_{p -1}^{(1)}$ is the (1,-1) spherical tensor component
of the dipole moment operator for ion $p$.
The rest of the calculation is very similar to the $\pi$ case.
The final result, analogous to Eq.~(\ref{crosssection10}) for the
$\pi$ case, is
\begin{equation}
\left\langle\frac{d\sigma^{(2)}}{d\Omega_{\rm out}}\right\rangle=
\frac{1}{8\pi}\sigma_0{\cal L}(\omega_{\rm in}-\omega_0),
\label{sigma1}
\end{equation}
which is independent of $\bbox{\hat{\rm k}}_{\rm out}$ and
shows no interference fringes.
The same result would have been obtained for any of the
other three initial states, since the absolute squares of the matrix elements
are the same.

For $\bbox{\hat\epsilon}_{\rm out}$=$\bbox{\hat\epsilon}_{\pi}$,
the matrix elements connecting the intermediate states to the final
states are
\begin{eqnarray}
\lefteqn{\langle\Psi_f|({\bf D}_1\cdot\bbox{\hat\epsilon}_{\pi})
e^{-i{\bf k}_{\rm out}\cdot {\bf R}_1}|\Psi_j\rangle}\nonumber\\
&=&\frac{-\cos\vartheta e^{i\varphi}}{\sqrt{2}}
\langle(^2S_{1/2},-1/2)_1|D_{1 -1}^{(1)}|(^2P_{1/2},+1/2)_1\rangle\nonumber\\
& &\times\langle\{n_{\rm ho}\}_f|e^{-i{\bf k_{\rm out}\cdot R}_1}|
\{n_{\rm ho}\}_j\rangle\\
&=&\frac{-\cos\vartheta e^{i\varphi}}{\sqrt{6}}(^2S_{1/2}\|D^{(1)}\|^2P_{1/2})
\langle\{n_{\rm ho}\}_f|e^{-i{\bf k}_{\rm out}\cdot {\bf R}_1}|
\{n_{\rm ho}\}_j\rangle.\nonumber
\end{eqnarray}
The final result is
\begin{equation}
\left\langle \frac{d\sigma^{(3)}}{d\Omega_{\rm out}}\right\rangle =
\frac{\cos^2\vartheta}{8\pi}\sigma_0{\cal L}(\omega_{\rm in}-\omega_0),
\label{sigma2}
\end{equation}
which shows no interference fringes.
The same result would have been obtained for any of the other initial states.

For the $\sigma$ case, Young's interference fringes are not observed
because only one of the two terms inside the absolute value
bars in  Eq.~(\ref{crosssection1}) is nonzero.
There is only one path for the photon, intersecting
the ion whose state is changed in the scattering process.

\subsection{Total cross section with or without polarization-selective
detection}

In Ref.~\cite{eichmann93}, a linear polarizer was sometimes placed
before the  photon detector.
For experimental convenience, the orientation of this polarizer was
fixed, while the input polarization could be varied.
To obtain the total cross section describing a given experimental
situation, we sum over all final atomic states and average over all
initial states.
For polarization-insensitive detection, we also sum over the polarizations
of the outgoing photon.

The cross section for polarization-insensitive detection is
\begin{eqnarray}
\left\langle \frac{d\sigma^{\rm unpol}}{d\Omega_{\rm out}}\right\rangle
&=&\left\langle\frac{d\sigma^{(1)}}{d\Omega_{\rm out}}\right\rangle
+2\left\langle\frac{d\sigma^{(2)}}{d\Omega_{\rm out}}\right\rangle
+2\left\langle\frac{d\sigma^{(3)}}{d\Omega_{\rm out}}\right\rangle\nonumber\\
&=&
\frac{\sigma_0}{4\pi}{\cal L}(\omega_{\rm in}-\omega_0)
\left\{1+\cos^2\vartheta\right.\nonumber\\
& & \left.\mbox{}+\sin^2\vartheta \left[1+\cos({\bf q\cdot d})
e^{-\frac{1}{2}\langle[{\bf q}\cdot({\bf u}_1-{\bf u}_2)]^2\rangle}
\right]\right\}.
\label{sigma-unpol}
\end{eqnarray}
The fringe visibility in this case cannot exceed 50\%.

The cross section for detection of light with polarization
$\bbox{\hat\epsilon}_{\pi}$ is
\begin{eqnarray}
\left\langle \frac{d\sigma^{(\pi)}}{d\Omega_{\rm out}}\right\rangle
&=&\left\langle\frac{d\sigma^{(1)}}{d\Omega_{\rm out}}\right\rangle
+2\left\langle\frac{d\sigma^{(3)}}{d\Omega_{\rm out}}\right\rangle\nonumber\\
&=&
\frac{\sigma_0}{4\pi}{\cal L}(\omega_{\rm in}-\omega_0)
\left\{\cos^2\vartheta\right.\nonumber\\
& &\left.\mbox{} +\sin^2\vartheta \left[1+\cos({\bf q\cdot d})
e^{-\frac{1}{2}\langle[{\bf q}\cdot({\bf u}_1-{\bf u}_2)]^2\rangle}
\right]\right\}.
\label{sigma-pi}
\end{eqnarray}
The fringe visibility in this case can approach 100\% in the $\vartheta=\pi/2$
plane if the Debye-Waller factor is close to 1.

The cross section for detection of light with polarization
$\bbox{\hat\epsilon}_{\sigma}$ is
\begin{equation}
\left\langle \frac{d\sigma^{(\sigma)}}{d\Omega_{\rm out}}\right\rangle
=2\left\langle\frac{d\sigma^{(2)}}{d\Omega_{\rm out}}\right\rangle
=\frac{\sigma_0}{4\pi}{\cal L}(\omega_{\rm in}-\omega_0),
\label{sigma-sigma}
\end{equation}
which is totally isotropic and shows no fringes.

\subsection{Which-path interpretation}

\begin{figure}[htbp]
\begin{center}
\subfigure[]{\includegraphics[clip,width=2.75in]{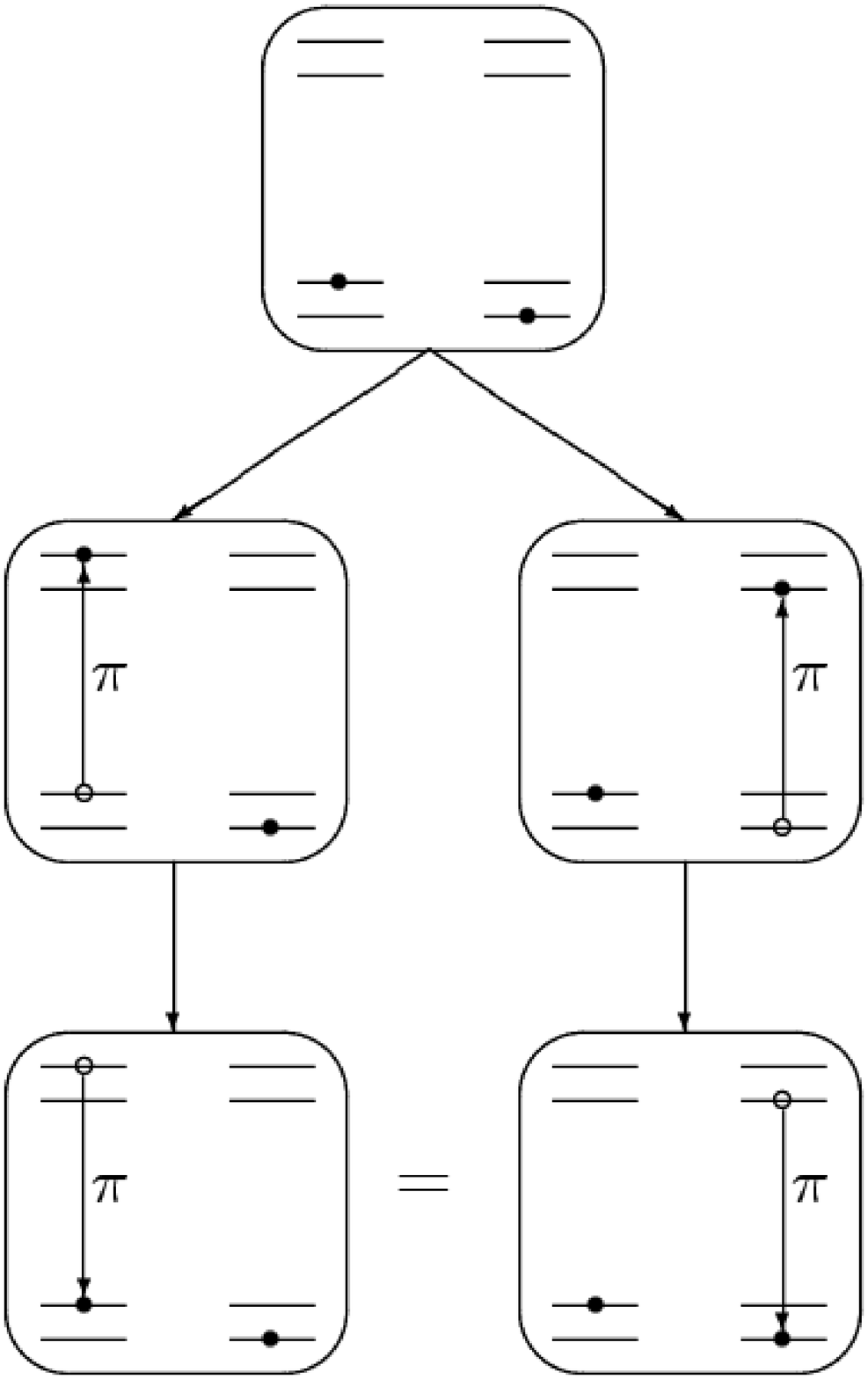}}
\hspace{.75in}
\subfigure[]{\includegraphics[clip,width=2.75in]{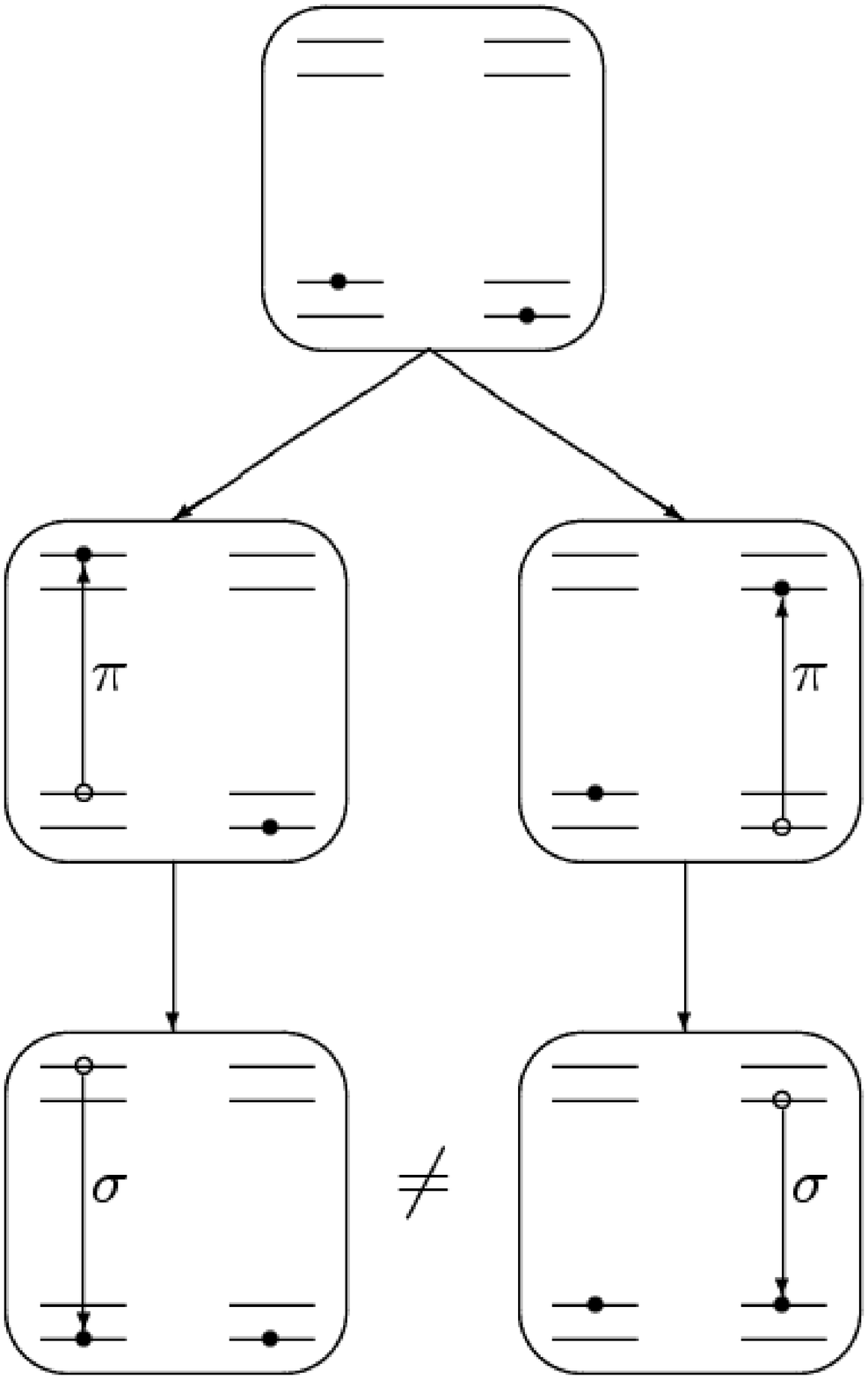}}
\end{center}
\caption{Each box represents the combined state of the two ions.
The ordering of energy levels is the same as in Fig.~\ref{levelfig}.
In (a) (the $\pi$ case),
one ion or the other undergoes a $\pi$ transition from the
ground to the excited state.
That ion undergoes a $\pi$ transition back to the ground state.
The two paths lead to the same final state of the two ions.
Hence, the probability amplitudes must be added, and interference
is possible.
In (b) (the $\sigma$ case), one ion or the other undergoes a
$\pi$ transition to the excited state, but
the excited ion undergoes a $\sigma$ transition to the ground state.
The two paths lead to different final states of the two ions.
Hence, there is no possibility of interference.
In order for interference to occur, it is {\em not}
necessary that the initial states
of the two ions be the same, only that the final combined states for
the two paths be the same.
}
\label{paths}
\end{figure}

The presence of interference fringes in the $\pi$ case and their
absence in the $\sigma$ case has a simple explanation in terms of the
possibility, in principle, of determining which of the two ions
scattered the photon.
Consider the sequence of transitions in Fig.~\ref{paths}(a),
representing the $\pi$ case.
Each box represents the combined state of the two ions.  Ion 1 is
represented by the diagram on the left side of a box
and ion 2 by that on the right.
The ordering of energy levels is the same as in Fig.~\ref{levelfig}.
For simplicity, we neglect the translational degrees of freedom,
which lead to the appearance of the Debye-Waller factor in
Eq.~(\ref{crosssection10}).
The system begins in the state
\begin{equation}
|\Psi_i\rangle = |(^2S_{1/2},+1/2)_1 (^2S_{1/2},-1/2)_2\rangle.
\label{init-state}
\end{equation}
One ion or the other absorbs a photon from the laser beam and
undergoes a $\pi$ transition to the excited state.
That ion emits a photon and
undergoes a $\pi$ transition back to the ground state.
The two paths, corresponding to either ion 1 or ion 2 scattering the
photon, lead to the same final state.
Therefore, the amplitudes for these two paths must be added, and
this leads to interference.
Since the final states of the ion are the same as the initial states,
it is not possible to determine which of the ions scattered the photon by
examining their states.

Now consider the sequence of transitions in Fig.~\ref{paths}(b),
representing the $\sigma$ case.
As in the previous case, one ion or the other absorbs a photon and undergoes
a $\pi$ transition to the excited state.
However, in this case, that ion undergoes a $\sigma$ transition when
it emits a photon and changes its $m_J$ quantum number.
The final states differ, depending on which of the ions scattered the
photon.
Hence, there is no interference between the two paths.
It would be possible to tell which ion scattered the photon by
examining the states of the ions before and after the scattering.

The preceding analysis is valid only in the limit of low laser intensity,
so that the probability of both ions being excited at the same time
is negligible, and stimulated emission can be neglected.
It is not necessary that the two ions be in the same quantum
state for interference to occur, only that the final combined
states for the two paths be indistinguishable.
For definiteness, a particular initial state [Eq.~(\ref{init-state})]
was chosen.
For each of the 3 other possible initial states, there is a process
like Fig.~\ref{paths}(a) in which the ions scatter a photon and
return to their original states
and one like Fig.~\ref{paths}(b)
in which one of them scatters a photon and changes its state.
Processes of the former type lead to interference; those of the
latter type do not.

\section{Comparison with experiment}

Figure \ref{fringes-2d} shows an image of the fringes observed for the
$\pi$ case, in which $\bbox{\hat{\epsilon}}_{\rm in}$ was perpendicular
to the $X$-$Z$ plane and the detector was sensitive only to light
polarized parallel to $\bbox{\hat{\epsilon}}_{\rm in}$.
The dark spots are due to stray reflections of the laser beams.
When $\bbox{\hat\epsilon}_{\rm in}$ was rotated by 90$^\circ$
without changing the polarizer in front of the detector
($\sigma$ case), the image showed no fringes.
The image data from a single ion, which shows no interference fringes,
were used to correct the data of Fig.~\ref{fringes-2d}
for a slowly spatially varying detection efficiency.
The data within the rectangle in Fig.~\ref{fringes-2d}
were summed along the vertical direction and divided by the
detection-efficiency function.

The normalized data points are shown in Fig.~\ref{fringe-fit} together
with a least-squares fit.
In this fit, as in Ref.~\cite{eichmann93},
the temperatures of the stretch and tilt modes
were assumed to have the ratio expected from theory \cite{itano82},
\begin{equation}
T_Z^{\rm rel}/T_X^{\rm rel}=
\{1+[3\cos^2(\Theta)]^{-1}\}/\{1+[3\sin^2(\Theta)]^{-1}\},
\end{equation}
and both temperatures were allowed to vary together in the fit.
The fringe visibility in the vicinity of the $X$-$Z$ plane is
insensitive to the temperature of the $Y$ motion, which is cooled indirectly
by coupling to the other modes.
The mean ion separation was calculated from knowledge of the trap parameters.
The dependence of Eq.~(\ref{crosssection10}) on the out-of-plane
angle $\Phi$ is small, and $\Phi$ was set to 0 in the fit.
The fitted value of $T_X^{\rm rel}$ was $1.08\pm 0.12$ mK,
or $0.92\pm 0.10$  times the Doppler-cooling limit.
The fringe visibility, extrapolated to $\phi=0$, would be
100\% if it followed Eq.~(\ref{crosssection10}).
The fitted value for this parameter was $(71 \pm 4)$\%.
The errors represent the standard deviations estimated from
the fit.
The maximum {\em observed} visibility, at the minimum value of
$\phi$ in Fig.~\ref{fringe-fit} is approximately 60\%.

There are several likely causes of the difference between the
observed and predicted values of the fringe visibility.
First, the theory was derived for the limit of low intensity.
The saturation parameter was measured to be $s=0.078\pm 0.025$
(see Appendix).
By itself, this would reduce the maximum visibility to
$(1+s)^{-1}\approx 93$\%, because the spectrum of the resonance fluorescence
in this polarization contains an incoherent part \cite{polder76}.
Other likely causes of reduced visibility are unequal laser
intensities at the two ions,
imperfect polarizers, stray background light,
and quantum jumps of one of the ions to a metastable state, leaving
only one ion fluorescing.
Each of these effects might reduce the visibility by a few percent.

\section{Discussion}

The fact that the resonance fluorescence from a two-level atom illuminated
by weak, monochromatic light is coherent with the applied field
was noted by Heitler \cite{heitler54}.
The spectrum of the resonance fluorescence for arbitrary applied intensities
was calculated by Mollow \cite{mollow69}.
In the limit of low applied intensity, the spectrum is monochromatic and
coherent with the applied field (a $\delta$ function).
At higher intensities, the coherent component decreases in amplitude,
and a component not coherent with the applied field and
having a width equal to the natural linewidth appears.
At very high intensities, the coherent component continues to decrease
in amplitude, and the incoherent component splits into three separate
Lorentzians.
The existence of a coherent component in the resonance fluorescence
of a single ion was confirmed directly by H\"offges {\em et al.}
by a heterodyne measurement \cite{hoffges97}.

Classically, we would expect the resonance fluorescence from
two two-level atoms at fixed positions, excited by the same monochromatic
field, to generate interference fringes having 100\%
visibility in the limit of low applied intensity, since the radiated fields
are coherent with each other.
At higher applied intensities, the visibility should decrease, since
more of the resonance fluorescence intensity belongs to the incoherent
component.
Quantum treatments for two two-level atoms
have been given by Richter \cite{richter91} and by
Kochan {\em et al.} \cite{kochan95}, who predict a visibility equal
to $(1+s)^{-1}$, where $s$ is the saturation parameter defined in
Ref.~\cite{cohen92}.
This is just the ratio of the intensity of the coherent component to
the total resonance fluorescence intensity for a single atom.

Polder and Schuurmans \cite{polder76} calculated the spectrum of the
resonance fluorescence of a $J=1/2$ to $J=1/2$ transition for a single atom.
The spectrum of the light having polarization $\bbox{\hat{\epsilon}}_\pi$
is like that for a two-level atom.
Hence, interference fringes would be expected in the
$\bbox{\hat{\epsilon}}_\pi$-polarized resonance fluorescence
from two such atoms for low applied intensity.
The spectrum of the light having polarization $\bbox{\hat{\epsilon}}_\sigma$
does not contain a $\delta$ function.
In the limit of low applied intensity, it is a Lorentzian having a width
approximately equal to the photon scattering rate, which can be much
less than the natural linewidth.
Even for applied intensities approaching $s=1$, the coherence length
is on the order of $c/\gamma$, where $\gamma$ is the spontaneous decay
rate of the excited state.
For the Hg$^+$ $6p\,^2P_{1/2}$ level, this is about 70 cm.
For interference fringes to exist, the radiation from the two atoms must be
{\em mutually} coherent.
Whether or not fringes should exist in the
$\bbox{\hat{\epsilon}}_\sigma$-polarized light from two atoms
is not immediately obvious from a classical analysis.
However, the perturbative quantum treatment of Sec.~V predicts
that there should be no interference, since there is only one
probability amplitude connecting the initial and final states.
The absence of interference in this case is fundamentally
a quantum effect, though one having more to do with the quantum
nature of the atom and the existence of degenerate, orthogonal
ground states, than with the quantum nature of the electromagnetic field.
Precisely the same point was made by Scully and Dr\"uhl
when they showed that interference fringes are not present in
the Raman radiation emitted by two three-level atoms having
a $\Lambda$ configuration \cite{scully82}.

Wong {\em et al.} \cite{wong97} calculated the interference of
resonance fluorescence from two four-level atoms having a level structure
like that of $^{198}$Hg$^+$.
Their analytic calculations are for a simpler geometry than
the one actually used by Eichmann {\em et al.} \cite{eichmann93}
and ignore the motion of the ions.
They do, however, include the effect of the decrease in visibility
due to the incoherent component of the resonance fluorescence,
which is not included in the perturbative calculation of Sec.~V.
The analytic calculations of Wong {\em et al.} and the present calculations
agree in the limits in which they are both valid, that is, for
low applied intensities and for no ion motion.
In particular, Wong {\em et al.} show that the fringe visibility can approach
100\% at low applied intensities, with polarization-selective detection.
Wong {\em et al.} also made Monte Carlo wavefunction simulations,
in which the motion of the ions was included classically,
and observed a decrease in visibility due to this effect.

Huang {\em et al.} \cite{huang96} calculated the effect of thermal motion
on the interference fringe visibility for two two-level atoms, each trapped
in a separate harmonic well.
They obtained an expression equivalent to Eq.~(1) of Eichmann
{\em et al.} \cite{eichmann93} for this model.
However, the treatment of Eichmann {\em et al.}, the details of which
are given in the present article, is more useful for the analysis of
the experiment of Ref.~\cite{eichmann93}, since it deals explicitly with
the actual normal mode structure of the two trapped ions.

Brewer has published a theory of interference in the light scattered
from two  four-level atoms \cite{brewer96b}.
One prediction of this theory is that the fringe visibility cannot
exceed 50\%, even with polarization-selective detection.
This contradicts the experimental results of Sec.~VI
shown in Fig.~\ref{fringe-fit}.
While the maximum visibility is about 60\%,
only slightly exceeding 50\%, {\em no} background has been
subtracted from the data, and there are several known sources of
decreased visibility, including thermal motion of the ions,
the incoherent component of the resonance fluorescence, and stray
scattered light.
The data were normalized by division by a slowly varying detection
sensitivity function, a process that cannot enhance the visibility.

The basic flaw in Brewer's argument can be seen in Eq.~(2) of
Ref.~\cite{brewer96b}, where he lists the basis states for the
two-atom system.
The states $|5\rangle$--$|8\rangle$ are the four states in which
both atoms are in the ground electronic state.
The states $|1\rangle$--$|4\rangle$ are linear combinations
of states in which one atom is in the ground state and one
is in the excited state.
However, most of the possible states of this type are missing,
apparently because of a false assumption that the allowed states must
have a particular kind of exchange symmetry.
For example, the intermediate superposition state shown in Fig.~\ref{paths}(a)
is, in his notation,
\begin{equation}
\frac{1}{\sqrt{2}}(|c_1b_2\rangle\pi_{nn,1}+|a_1d_2\rangle\pi_{nn,2}),
\end{equation}
and is not contained in the list.
The neglect of these basis states leads to the neglect of processes like
that of Fig.~\ref{paths}(a), in which the two atoms are initially in
different $m_J$ states.
Thus, he reaches the false conclusion that the two atoms must initially
be in the same $m_J$ state in order for interference to occur.
Since he misses half of the processes that lead to interference,
he predicts a maximum visibility, with polarization-sensitive detection,
of 50\% rather than 100\%.

We conclude with some remarks regarding the principle of complementarity.
Wave and particle properties of light
are complementary and hence cannot be observed at the same time.
If it is possible to determine which atom scattered the photon, the
interference fringes must vanish.
Feynman's thought experiments, in which various methods of determining
the path of an electron through a two-slit Young's interferometer
lead to the destruction of interference fringes due
to a random momentum transfer
are often quoted [Ref.~\cite{feynman65}, pp. 1-6--1-11].
However, in Ch.~3 of the same textbook, Feynman emphasizes the
seemingly more fundamental viewpoint that interference is present
only if there exist different indistinguishable ways to go from a
given initial state to the {\em same} final state.
His example of the scattering of neutrons from a crystal is very
similar to the experiment of Eichmann {\em et al.}
If the nuclei of the atoms in the crystal have a nonzero spin, the
angular distribution of scattered neutrons is the sum of a featureless
background and some sharp diffraction peaks.
The sharp diffraction peaks are associated with neutrons which do not
change their spin orientations in the scattering.
The featureless background is associated with neutrons whose spins
change their orientations in the scattering.
In this case, there must also be a change in the spin orientation of one
of the nuclei in the crystal.
It would be possible in principle to determine the nucleus which
scattered the neutron, so there is no interference.
The position-momentum indeterminacy relations play no essential role
in the presence or absence of interference in this case,
just as in the experiment of Eichmann {\em et al.}

\section*{Acknowledgments}
This work was supported by the Office of Naval Research.
U.\ E.\ acknowledges financial support from the Deutsche
Forschungsgemeinschaft.
Dr.\ J.\ M.\ Gilligan assisted in the early stages of the experiment
and suggested the method that was used for the polarization-selective
detection.

\appendix
\section*{Calibration of the saturation parameter}

For the case where an electric-dipole transition  between
a $^2S_{1/2}$ ground state and a $^2P_{1/2}$ excited state
is excited by linearly polarized light,
we define the saturation parameter $s$ similarly to the way in which
it is defined for a two-level system \cite{cohen92}.
The magnetic field is assumed to be small, and the quantization axis
for the ion is along the electric field.
We define
\begin{equation}
s=\frac{\Omega_1^2/2}{(\omega_0-\omega_{\rm in})^2+(\gamma/2)^2},
\end{equation}
where $\Omega_1=6^{-1/2}|{\cal E}_0(^2S_{1/2}\|D^{(1)}\|^2P_{1/2})|\hbar^{-1}$
is the Rabi frequency, and the other terms have been defined previously.
In order for the perturbative analysis of Sec.~V to be valid,
we must have $s\ll 1$.
In the case of Hg$^+$, the $^2P_{1/2}$ state has a small (approximately
10$^{-7}$ ) probability of decaying to the metastable $^2D_{3/2}$ state,
which decays either directly to the ground state or to the metastable
$^2D_{5/2}$ state, which decays to the ground state.
The 194.2 nm fluorescence intensity from a single ion is bistable,
since it has a steady level when the ion is cycling between
the $^2S_{1/2}$ and $^2P_{1/2}$ states and
vanishes when the ion drops to a metastable state.
The fractional population of the $^2P_{1/2}$ state, summed over both $m_J$
values, is $s/[2(1+s)]$  while the ion is cycling between the
$^2S_{1/2}$  and $^2P_{1/2}$ states.
The quantum jump statistics have been discussed in several previous
articles \cite{itano87,itano87b,itano88}.
For a single ion, we define $p_{\rm on}$ to be the fraction of the time
that the ion is cycling between the $^2S_{1/2}$ and $^2P_{1/2}$ states,
and $p_{\rm off}$=$(1-p_{\rm on})$ to be the fraction of the time
that it spends in either of the metastable states.
It can be shown, from the steady-state solutions of the differential
equations for the populations [Eqs.~(2a)--(2c) of Ref.~\cite{itano88}],
that $s$ is related to the ratio $p_{\rm off}/p_{\rm on}$
according to
\begin{equation}
\frac{1}{2} \frac{s}{(1+s)} = \frac{\gamma_1\gamma_2(p_{\rm off}/p_{\rm on})}
{\gamma_3(\gamma_2+f_2\gamma_1)}\approx 0.36\;\frac{p_{\rm off}}{p_{\rm on}},
\label{s_calib}
\end{equation}
where the parameters $\gamma_1$, $\gamma_2$, $\gamma_3$, and $f_2$
have been measured \cite{itano87},
and the uncertainty in the coefficient (0.36) is about 30\%, due mostly to
the uncertainty in $\gamma_3$.

\begin{figure}[htbp]
\begin{center}
\includegraphics[clip,width=3in]{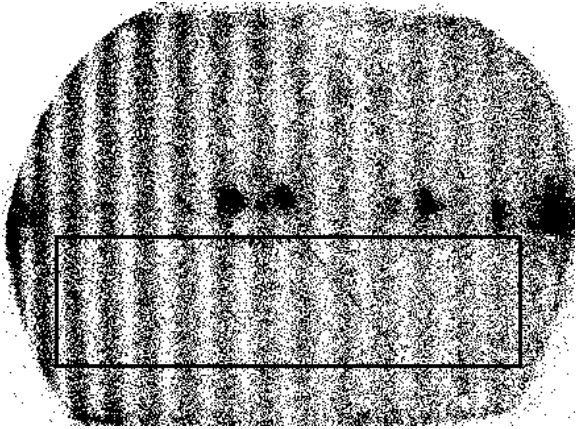}
\end{center}
\caption{Experimental fringe data for the case in which the
the detected light is polarized in the same direction
as the incoming light ($\pi$ case).
The ion separation $d=4.17$ $\mu$m.
The angle $\phi$ (the deviation from the forward scattering
direction) increases to the right.
The decrease in visibility with increasing $\phi$ is due to thermal
motion of the ions.
The dark spots are due to stray reflections of the laser beams.
The data within the rectangle were summed along the vertical direction
and least-squares fitted.
}
\label{fringes-2d}
\end{figure}

\begin{figure}[hbp]
\begin{center}
\includegraphics[clip,width=3in]{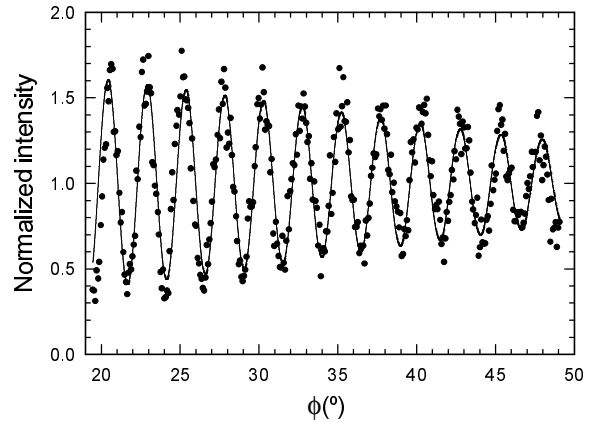}
\end{center}
\caption{Experimental fringe data (dots) from the image of
Fig.~\ref{fringes-2d} and a least-squares fit (line) to the sum of
the theoretical intensity [Eq.~(\ref{crosssection10})] and a constant
background.
The fitted temperature is approximately equal to the Doppler-cooling
limit.
}
\label{fringe-fit}
\end{figure}

\begin{figure}[htbp]
\begin{center}
\includegraphics[clip,width=3in]{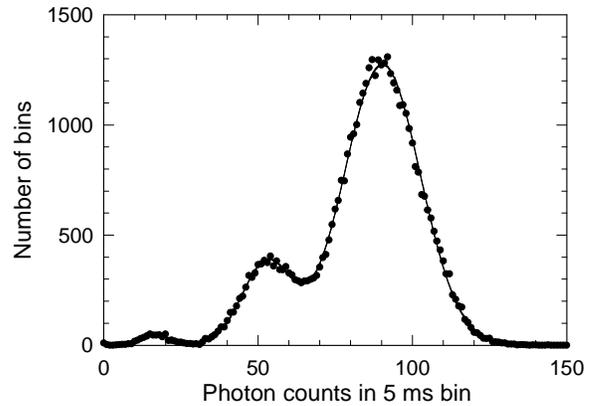}
\end{center}
\caption
{Plot of the probability distribution of the fluorescence
intensity for two ions, used to determine the saturation parameter $s$.
The horizontal axis corresponds to the number of photons  counted
in a 5 ms interval.
The vertical axis corresponds to the number of 5 ms intervals in which
a given number of photons was counted.
This was measured simultaneously with the
interference fringes shown in Fig.~\ref{fringes-2d}
The curve is a least-squares fit to a sum of three Gaussians.
The areas under the Gaussians, from left to right,
are proportional to the probabilities
that 2, 1, or 0 of the ions are in a metastable state.
Higher values of $s$ correspond to higher populations in the metastable
states.}
\label{tristable}
\end{figure}

For two ions, the fluorescence will be tristable, since 0, 1, or 2
ions may be in a metastable state.
During an interference fringe measurement, the number of
photons detected in each successive period of a few milliseconds was recorded.
Figure~\ref{tristable} shows a plot of the probability distribution
of the 5 ms photon counts during the measurement of Fig.~\ref{fringes-2d}.
The three peaks correspond, from left to right,
to 2, 1, or 0 ions being in a metastable state.
The leftmost peak corresponds to the signal from stray background
light, since there is no fluorescence from the ions.
The curve is a least-squares fit to a sum of three Gaussians.
The areas under the peaks should be in the ratio
$p_{\rm off}^2$:$2p_{\rm off}p_{\rm on}$:$p_{\rm on}^2$.
The ratios of the areas obtained from the fit are
0.011:0.160:0.828, so $p_{\rm off}/p_{\rm on}$=$0.10\pm 0.01$,
and, from Eq.~(\ref{s_calib}), $s$=$0.078\pm 0.025$,
so the perturbative analysis should be a good approximation.

During this measurement period,
the interference fringe detection was gated off for 5 ms
if the number of photons detected in the previous 5 ms was less than 80.
This helped to prevent loss of the fringe visibility due to
background from single-ion fluorescence, which would have no
interference fringes.

\end{multicols}

\twocolumn

\onecolumn

\twocolumn

\end{document}